\documentclass{elsart}
\usepackage{graphicx}
\usepackage{epsfig}
\usepackage{amssymb}
\usepackage{color}

\begin{document}
 \begin{frontmatter}

 \title{Multilayer Adsorption of Interacting Polyatomics on Heterogeneous Surfaces}

\author[label1,label2]{F. O. S\'anchez-Varretti},
\author[label1,label2]{G. D. Garc\'{\i}a},
\author[label1]{A. J. Ramirez-Pastor\corauthref{cor1}}

\address[label1]{Dpto. de F\'{\i}sica, Instituto de F\'{\i}sica Aplicada, Universidad Nacional de San Luis - CONICET, Chacabuco 917, 5700 San Luis, Argentina.}
\address[label2]{Universidad Tecnol\'ogica Nacional, Regional San Rafael, Gral. Urquiza 314, 5600, San Rafael, Mendoza, Argentina. }
\thanks[cor1]{Corresponding author. Fax +54-2652-430224, E-mail: antorami@unsl.edu.ar}

\begin{abstract}
In the present work we introduce a generalized lattice-gas model
to study the multilayer adsorption of interacting polyatomics on
heterogeneous surfaces. Using an approximation in the spirit of
the well-known Brunauer--Emmet--Teller (BET) model, a new
theoretical isotherm is obtained in one- and two-dimensional
lattices and compared with Monte Carlo simulation. In addition, we
use the BET approach to analyze these isotherms and to estimate
the monolayer volume. In all cases, we found that the use of the
BET equation leads to an underestimate of the true monolayer
capacity. However, significant compensation effects were observed
for heterogeneous surfaces and attractive lateral interactions.
\end{abstract}

\end{frontmatter}

\newpage


\section{Introduction}

The theoretical description of adsorption is a long-standing
complex problem in surface science that presently does not have a
general solution
\cite{Clark1970,Steele1974,GreggSing1991,Adamson1990}.

In 1918 Langmuir derived the monolayer adsorption isotherm
kinetically for gas monoatomic molecules adsorbed on the
homogeneous surface of adsorbents without attractions among the
adsorbed molecules \cite{Langmuir1918}. Later, some theories have
been proposed to describe multilayer adsorption in equilibrium
\cite{BET1938,Frenkel1955,Halsey1948,Hill1952,McMillan1951,McMillan1951b,Dreyfus,Cortes1987,Payne1996,Casal2002,Grabowskyi2002,Rudzinski1992}.
Among them, the one of Brunauer-Emmett-Teller (BET) \cite{BET1938}
and the one of Frenkel-Halsey-Hill
\cite{Frenkel1955,Halsey1948,Hill1952} are the simplest which
provide the basis to construct more elaborate approaches. Those
more elaborate analytic approaches take into account lateral
interactions between the admolecules, differences between the
energy of the first and upper layers, surface energetic
heterogeneity and so forth. These leading models have played an
important role in the characterization of solid surfaces by means
of gas adsorption.

A more accurate description of multilayer adsorption should
account for the fact that, in practice, most adsorbates are
polyatomic. Even the simplest nonspherical molecules such as $N_2$
and $O_2$ may adsorb on more than one site depending on the
surface structure
\cite{Rudzinski1992,Zeppenfeld1997,Mouristen1982,Ferry1996,He1992,Panella1994,Meixner1992}.
 This effect, called multisite-occupancy
adsorption, introduces a new complexity to the adsorption theory.

The difficulty in the analysis of the multisite statistics is
mainly associated with three factors which differentiate the
$k$-mers statistics from the usual single-particle statistics.
Namely, $(i)$ no statistical equivalence exists between particles
and vacancies; $(ii)$ the occupation of a given lattice site
ensures that at least one of its nearest-neighbor sites is also
occupied; and $(iii)$ an isolated vacancy cannot serve to
determine whether that site can ever become occupied.

However, several attempts were done in the past in order to solve
the $k$-mers problem. Among them, Onsager \cite{ONSAGER}, Zimm
\cite{ZIMM} and Isihara \cite{ISIHARA} made important
contributions to the understanding of the statistics of rigid rods
in dilute solution. The FH theory, due independently to Flory
\cite{FLORY} and to Huggins \cite{HUGGI}, has overcome the
restriction to dilute solution by means of a lattice calculation.
The approach is a direct generalization of the theory of binary
liquids in two dimensions or polymer molecules diluted in a
monomeric solvent. It is worth mentioning that, in the framework
of the lattice-gas approach, the adsorption of $k$-mers on
homogeneous surfaces is an isomorphous problem to the binary
solutions of polymer-monomeric solvent.

More recently, two new theories to describe adsorption with
multisite occupancy have been introduced. In the first,
Ramirez-Pastor et al. \cite{PRB3,LANG5,LANG9} presented a model to
study the adsorption of linear adsorbates on homogeneous surfaces.
The model is based on exact forms of the thermodynamic functions
of linear adsorbates in one dimension and its generalization to
higher dimensions. In the second, which is called the fractional
statistical theory of the adsorption of polyatomics (FSTA), the
configuration of the molecule in the adsorbed state is
incorporated as a model parameter \cite{PRL1,IJMP}.

The treatments in Refs.
\cite{ONSAGER,ZIMM,ISIHARA,FLORY,HUGGI,PRB3,LANG5,LANG9,PRL1,IJMP}
are limited in their application because they are valid only for
monolayer adsorption. There are few studies that take into account
multisite occupancy at the multilayer regime
\cite{Aranovich1995,Aranovich1997,Riccardo2002,Roma2005,Garcia2009,Sanchez2009,Nikitas1996}.
Aranovich and Donohue \cite{Aranovich1995,Aranovich1997} presented
a multilayer adsorption isotherm that should be capable to include
multisite occupancy. Later, a closed exact solution for the
multilayer adsorption isotherm of dimers was reported in Ref.
\cite{Riccardo2002}. It demonstrates that the determinations of
surface areas and adsorption energies from polyatomic adsorbate
adsorption (also called $k$-mers) may be misestimated, if this
polyatomic character is not properly incorporated in the
thermodynamic functions from which experiments are interpreted.
This treatment has recently been extended \cite{Roma2005} to
include the two-dimensional nature of the substrate.

There are another two important physical facts which have not been
sufficiently studied: $1)$ the effect of the lateral interactions
between the ad-molecules and $2)$ the effect of surface
heterogeneity in presence of multisite and multilayer adsorption.
In the first case, a recent paper \cite{Garcia2009} extends the
BET equation to include nearest-neighbor lateral interactions
between the molecules adsorbed in the first layer. Following the
configuration-counting procedure of the Bragg-Williams approach
and the quasi-chemical approximation, Ref. \cite{Garcia2009} gives
us a simple statistical mechanical framework for studying
multilayer adsorption of interacting polyatomics. In the second
case, multilayer adsorption of polyatomics on heterogeneous
surfaces is a topic of interest in surface science since it
comprises most features present in experimental situations.
Surfaces generally present inhomogeneities due to irregular
arrangement of surface and bulk atoms, the presence of various
chemical species, etc. In this context, a recent job shows how the
monolayer volume predicted by BET equation differs from its real
value when considering both the adsorbate size and the surface
topography \cite{Sanchez2009}.

On the other hand, combined effects coming from lateral
interactions, multisite occupancy and surface heterogeneity have
been analyzed in the interesting paper by Nikitas
\cite{Nikitas1996}. In Ref. \cite{Nikitas1996}, the author
concludes that: $(i)$ one can obtain an underestimation of the
true monolayer capacity of the order of $25 \%$ when the adsorbate
occupies more than one lattice site, and this underestimation will
become worse if the effect of the multisite occupancy is coupled
with heterogeneity effects and $(ii)$ the attractive interactions
in the gas adsorption lead always to a weak overestimation of the
monolayer volume.

The previously discussed issues can be outlined in Fig. 1, where
dimensions of complexity are sketched. In each axis we see a
particular characteristic of the adsorption process that can (or
cannot) be present in a particular model. Usually, the more
features a model has,  the more is its generality; and more
complex the process it describes. In the multilayer regime, we can
see models that include adsorption with lateral interactions, but
they do not allow to multisite occupancy, or incorporate surface
energetic heterogeneity, but they overlook lateral interactions in
the adsorbed layer, etc. With Fig. 1 we try to show that, at
present, some models venture in some of these dimensions of
complexity but, the vast majority, lack the complete set of
complexities.  It is also clear that the farther away we are from
the origin the more general (and usually the more complex) the
model is. Even more, the upper plane (including multisite
occupancy), is a set of models developed in the last decade that
represent a leap forward in the lattice-gas theory.

In this work we will attempt to take one step further away from
the origin, into a more general and encompassing scheme,
presenting a model that considers the totality of these dimensions
of complexity (multilayer adsorption, lateral interactions,
multisite occupancy, energetic heterogeneity and surface
topography). For this purpose, a theoretical formalism is
presented based upon the analytical expression of the adsorption
isotherm with lateral interaction weighted by the characteristic
length of the surface heterogeneous. In addition, Monte Carlo (MC)
simulations are performed in order to test the validity of the
theoretical model. The new theoretical scheme allows us $(1)$ to
obtain an accurate approximation for multilayer adsorption on
two-dimensional substrates accounting multisite occupancy, lateral
interactions, energetic heterogeneity and surface topography, and
$(2)$ to provide a simple model from which experiments may be
reinterpreted.

The present work is organized as follows. In Section II, we
present the theoretical formalism. Section III is devoted to
describing the simulation scheme. The analysis of the results and
discussion are given in Section IV. Finally, the conclusions are
drawn in Section V.


\section{Model and theory}

In this section, we will present the model and make a revision of
some existing theories describing the adsorption process in the
framework of the lattice-gas approximation. We will focus on those
contributions that give contextualized understanding of the work
that we will present.

\subsection{Model}

The substrate is modeled by a regular lattice of $M$ sites with
periodic boundary conditions, where the adsorption energy of the
first layer $\varepsilon^f_i$ depends on each site $i$ of the
surface and the adsorbate is represented by $k$-mers (linear
particles that have $k$ identical units). A $k$-mer adsorbed
occupies $k$ sites of the lattice and can arrange in many
configurations. This property is called multisite-occupancy
adsorption.

On the other hand, for higher layers, the adsorption of a $k$-mer
is exactly onto an already absorbed one, with an adsorption energy
of $k \varepsilon$. Thus, the monolayer structure reproduces in
the remaining layers. The mechanism used to describe the
adsorption in the multilayer regime mimics the phenomenon called
pseudomorphism. This was observed, by using low-energy electron
diffraction technique \cite{Firment1976,Firment1978,Somorjai1979},
in the case of adsorption of straight chain saturated hydrocarbon
molecules on metallic surfaces. However, recent synchrotron X-ray
scattering measurements show a different and more complex growth
process in these systems \cite{Wu2001}. Nevertheless, for
simplicity, here we study a variant of the model that maintains
the same mechanism to describe the adsorption in the higher
layers.  Because we analyze isotherms in the low coverage regime,
we expect that this choice will not affect our results.

Finally, attractive and repulsive lateral interactions are
considered in the first layer and horizontal interactions are
ignored in higher layers. In this sense, although the contribution
from the secondary adsorption can already be significant, the
density of the molecules in the second and higher adlayers is
expected to be much lower than that in the first adsorbed layer
[BET equation can be applied at coverage not greatly exceeding
(statistically) monolayer coverage]. Therefore, it seems to be
satisfactorily enough to take into account only the interactions
between the primarily adsorbed molecules \cite{Hill}.

Under these considerations we can write the Hamiltonian as:
\begin{equation}
H=\sum _{i=1} ^M \varepsilon^f_i \sigma_i + k \varepsilon (N-N_1)+
w \sum _{\langle i,j \rangle} \sigma_i \sigma_j - w N_1 (k-1),
\end{equation}
where the first term of the right-hand side (RHS) represents the
adsorption energy of the $N_1$ $k$-mers adsorbed in the first
layer (adjacent to de adsorbent) and the second term is the energy
of the $(N-N_1)$ $k$-mers adsorbed on top of the first layer
(second layer, third layer, and so on). The third and fourth terms
correspond to the lateral interaction energy, where $w$ is the
interaction energy between two nearest-neighbor (NN) units
belonging to different $k$-mers adsorbed in the first layer (we
use $w>0$ for repulsive and $w<0$ for attractive interactions);
$\sigma_i$ is the occupation variable which can take the values 0
if the site $i$ is empty or 1 if the site $i$ is occupied and
$\langle i,j \rangle$ represents pairs of NN sites. Since the
summation in the third term overestimates the total energy by
including $N_1(k-1)$ bonds belonging to the $N_1$ adsorbed
$k$-mers, the fourth term subtracts this exceeding energy.

Our study will be restricted to the class of lattice-gas models in
which the substrate is a regular array of individual adsorption
sites where molecules can be deposited (also in a discrete
manner). In the following sections we will review some analytical
deductions of those models. A more in depth treatment is carried
out in the original publications.

\subsection{Multilayer adsorption of non-interacting polyatomics on homogeneous surfaces}

One way to begin our analysis is to review the behavior of a
system of identical particles, which will be adsorbed in a regular
lattice of $M$ identical sites of adsorption \cite{Roma2005}. The
supposition is that, like in the BET model, adsorption is done on
the surface of the solid or onto an already adsorbed particle. In
this case, in stead of adsorbing spherically symmetric ad-atoms,
we will use entities (that we will call $k$-mers) that occupy $k$
consecutive lattice sites.

Therefore, we will only have two possible unit-adsorption
processes: $(1)$ a $k$-mer occupies $k$ consecutive empty surface
sites; and $(2)$ a $k$-mer adsorbs on top of an already adsorbed
$k$-mer. This adsorptive process will form columns of $k$-mers on
the solid. We must notice that a unit-adsorption where a $k$-mer
adsorbs on top of two already adsorbed $k$-mers is prohibited.

In the case of the unit-desorption processes, we will only be able
to desorb a $k$-mer that is on the top of one $k$-mer column or a
$k$-mer that is adsorbed on the solid surface, but has no other
$k$-mers adsorbed on top of it.

Under these conditions, the grand partition function of the system
is:
\begin{equation}
\Xi = \sum ^{n_{max}}_{n=0} \Omega _k(n,M) \xi^n,
 \label{eq:newtheoricalmulticapa}
\end{equation}
where $n_{max}(=M/k)$ is the maximum number of columns that can be
formed, $\Omega_k(n,M)$ it is the total number of distinguishable
configurations of $n$ columns in $M$ sites and $\xi$ is the grand
partition function of a unique column of $k$-mers that has at
least one $k$-mer in the first layer.

On the other hand, we have that the grand partition function for
the monolayer ($\Xi_{1}$) is:
\begin{equation}
 \Xi _{1} = \sum ^{n_{max}} _{n=0} \Omega _k (n,M) \lambda_{1}
 ^{n},
 \label{eq:newtheoricalmonocapa}
\end{equation}
being $n$, in this case, the number of $k$-mers adsorbed on the
surface of the solid (first layer), and $\lambda _{1}$ the
fugacity of the monolayer. $\Omega _k (n,M)$ is still the number
of possible configurations with $n$ $k$-mers in $M$ sites. This
quantity must be equal in Eqs. (\ref{eq:newtheoricalmulticapa})
and (\ref{eq:newtheoricalmonocapa}).

If now we compare Eqs. (\ref{eq:newtheoricalmulticapa}) and
(\ref{eq:newtheoricalmonocapa}) we can observe that they have a
similar form. This allows us to write:
\begin{equation}
\left( \frac{\partial \Xi }{\partial \xi } \right) _{M,T} =
\left(\frac{\partial \Xi _{1} }{\partial \lambda _{1}}\right)
_{M,T}.
\end{equation}
If, in addition, inspired by this similarity, we propose the
following ansatz:

\begin{equation}
\lambda _{1}= \xi.
\end{equation}
Then
\begin{equation}
\lambda_{1} = \xi = \frac{c p/p_0}{1-p/p_0} \Rightarrow
\frac{p}{p_0} = \frac{1}{1+c/\lambda_{1}}, \label{eq:4NT}
\end{equation}
where $c=q_1/q=\exp[-\beta k(\varepsilon^f-\varepsilon)]$ is the
ratio between the partition function of a particle in the first
layer and a particle in any other layer [$\beta=(k_BT)^{-1}$,
being $k_B$ the Boltzmann constant]. And, then, we can write the
monolayer coverage as:

\begin{eqnarray}
\theta_{1} &= & \frac{k \bar n}{M} =\frac{k}{M} \lambda_{1} \left(
\frac{\partial \ln \Xi _{1}}{\partial \lambda_{1}}
\right)_{M,T} \\
 & = & \frac{k}{M} \xi \left( \frac{\partial \ln \Xi}{\partial \xi}
 \right)_{M,T}.
\end{eqnarray}

Now, the total coverage ($\theta$) can be written in terms of the
coverage of the monolayer ($\theta_{1}$):

\begin{equation}
\theta = \frac{\theta_{1}}{1-p/p_0}. \label{eq:7NTa}
\end{equation}

The theoretical procedure in Eqs. (\ref{eq:4NT})-(\ref{eq:7NTa})
provides the isotherm in the multilayer regime from the isotherm
in the monolayer regime. In fact:

(1) By using $\theta _{1}$ as a parameter ($0 \leq \theta _{1}
\leq 1$), the relative pressure is obtained by using Eq.
(\ref{eq:4NT}). This calculation requires the knowledge of an
analytical expression for the monolayer adsorption isotherm.

(2) The values of $\theta _{1}$ and $p/p_0$ are introduced in Eq.
(\ref{eq:7NTa}) and the total coverage is obtained. The items (1)
and (2) are summarized in the following scheme:
$$\theta _{1} + \lambda _{1}(\theta _{1})+ \textrm{Eq.} \ \ (\ref{eq:4NT}) \rightarrow p/p_0$$
$$ \Rightarrow \theta _{1} + p/p_0 + \textrm{Eq.} \ \ (\ref{eq:7NTa}) \rightarrow
\theta.
$$

Following the previous scheme, it is possible to obtain the exact
multilayer isotherm for $k$-mers in 1D homogeneous surfaces. In
fact, in Ref. \cite{Roma2005} the expression for the monolayer
coverage is proved to be:
\begin{equation}
p/p_0=\frac{\theta_{1} \left[1- \frac{(k-1)}{k} \theta _{1}
\right]^{k-1}}{kc(1-\theta _{1})^k+\left[1-\frac{(k-1)}{k} \theta
_{1}\right]^{k-1}}. \label{eq:21NT}
\end{equation}
Equations (\ref{eq:7NTa}) and (\ref{eq:21NT}) represent the exact
solution of the 1D model and, as it is expected, retrieve the BET
equation for the case $k=1$.

Also, the previous scheme can be used to obtain an accurate
approximation for multilayer adsorption on 2D substrates
accounting multisite occupancy. In this case, the semi-empirical
monolayer adsorption isotherm \cite{IJMP,Roma2006} can be used
\begin{equation}
\frac{p}{p_0} = \frac{ \theta_1 \left[ 1- \frac{\left( k-1
\right)}{k} \theta_1 \right]^{\left(k-1 \right) \theta_1} \left[
1- \frac{2 \left( k-1 \right)}{\zeta k} \theta_1
\right]^{\left(k-1 \right) \left( 1- \theta_1 \right)}}{z k c
\left( 1-\theta_1\right)^k +  \theta_1 \left[ 1- \frac{\left( k-1
\right)}{k} \theta_1 \right]^{\left(k-1 \right) \theta_1} \left[
1- \frac{2 \left( k-1 \right)}{\zeta k} \theta_1
\right]^{\left(k-1 \right) \left( 1- \theta_1 \right)}},
\label{pre2D}
\end{equation}
where $\zeta$ is the connectivity of the lattice and $z$
represents the number of available configurations (per lattice
site) for a linear $k$-mer at zero coverage. Thus, $z=1$ for $k=1$
and $z=\zeta/2$ for $k \geq 2$.

Note that, for $\zeta = 2$, Eq. (\ref{pre2D}) is identical to the
Eq. (\ref{eq:21NT}).  Therefore, Eqs. (\ref{eq:7NTa}) and
(\ref{pre2D}) represent the general solution of the problem of
multilayer adsorption in homogeneous surfaces with multisite
occupancy.

In the following section our aim will be to generalize this
expression to include the lateral interactions following the
methodology used in Ref. \cite{Garcia2009}. Clearly, the
complexity of the isotherm is greatly increased, but the
expression is still a manageable one and in the case of zero
interactions the previous case is retrieved.

\subsection{Multilayer adsorption of interacting polyatomics on homogeneous surfaces}

Due to the complexity introduced in the analytical expressions
because of the lateral interactions is that approximations are
used to deal with this feature of the adsorption process. The most
commonly used approximations are the {\em mean-field
approximation} (MFA) \cite{Hill,Davila06} and the {\em
quasi-chemical approximation} (QCA) \cite{Hill,Davila06}.

As shown in previous work\cite{Davila06}, the
configuration-counting procedure \footnote{The important
assumption in this method is that pairs of NN sites are treated as
if they were independent of each other (this assumption is, of
course, not true, because the pairs overlap \cite{Hill}).} of the
QCA allows us to obtain an approximation that is significantly
better than the MFA for polyatomics. Based on this finding, we
restrict the rest of our discussion to the estimates obtained
under QCA.

In order to apply the theoretical scheme described in previous
section, we start with the monolayer adsorption isotherm of
interacting $k$-mers adsorbed on a lattice of connectivity $\zeta$
obtained from the formalism of QCA \cite{Davila06},
\begin{equation}
\lambda_1=\left[\frac{\theta_1 \exp {\left( \beta w z /2
\right)}}{k \ \eta(\zeta,k) \left( \frac{2}{\zeta}
\right)^{2(k-1)}}\right] \left[
\frac{(1-\theta_1)^{k(\zeta-1)}\left[ k - (k-1) \theta_1
\right]^{k-1}
 \left[\frac{z \theta_1}{2k}- \alpha \right]^{z/2}}{\left[\frac{\zeta k}{2}-(k-1)\theta_1) \right]^{k-1} \left[\frac{\zeta}{2}(1-\theta_1)- \alpha \right]^{k \zeta /2} \left(\frac{z \theta_1}{\zeta k}
 \right)^{z}}\right],
 \label{mu}
\end{equation}
where
\begin{equation}
z=\left[ 2(\zeta-1)+(k-2)(\zeta-2)\right] \label{lam},
\end{equation}
\begin{equation}
\alpha = \frac{z \zeta}{2k} \frac{\theta_1
(1-\theta_1)}{\left[\frac{\zeta}{2}-\left(\frac{k-1}{k}
\right)\theta_1 + b\right]},    \label{alfa}
\end{equation}
\begin{equation}
b= \left\{ \left[\frac{\zeta}{2}-\left(\frac{k-1}{k}
\right)\theta_1 \right]^2  - \frac{z \zeta}{k} A \theta_1
(1-\theta_1) \right\}^{1/2},    \label{b}
\end{equation}
and
\begin{equation}
A= 1-\exp(-\beta w).
\end{equation}

Replacing Eq. (\ref{mu}) into Eq. (\ref{eq:4NT}), we obtain
\begin{equation}
\left({p \over {p_o}}\right)^{-1}  =  1+\frac{c k \eta(\zeta,k)
\left( \frac{2}{\zeta} \right)^{2(k-1)} \left[\frac{\zeta
k}{2}-(k-1)\theta_1) \right]^{k-1}
\left[\frac{\zeta}{2}(1-\theta_1)- \alpha \right]^{k \zeta /2}
\left(\frac{z \theta_1}{\zeta k} \right)^{z}} {\theta_1 \exp
{\left( \beta w z /2 \right)}(1-\theta_1)^{k(\zeta-1)}\left[ k -
(k-1) \theta_1 \right]^{k-1}
 \left[\frac{z \theta_1}{2k}- \alpha \right]^{z/2}}
\label{pqca}.
\end{equation}
Eqs. (\ref{eq:7NTa}) and (\ref{pqca}) represent the solution
describing the multilayer adsorption of interacting $k$-mers on
homogeneous surfaces in the framework of the QCA. This method is
presented in more depth in \cite{Garcia2009}.

Now we can see that we have moved upward in the plane of
complexity and further more included interactions. The next
section revises the role of surface heterogeneity and lateral
interactions.

\subsection{Multilayer adsorption of interacting polyatomics on heterogeneous surfaces}

In the two previous sections, we obtained the multilayer isotherm
from the monolayer isotherm. It is possible demonstrate that this
formalism still holds for interacting $k$-mers and a given surface
heterogeneity. However, this strategy leads to a complex solution
that is not useful for practical purposes. To build a simpler
function (easier to analyze), we have chosen to approximate the
multilayer heterogeneous isotherm by a weighted sum of multilayer
homogeneous isotherms.

The heterogeneous surface is modeled by two kinds of adsorption
sites in the first layer (bivariate surface): strong sites with
adsorption energy $\varepsilon^f_1$ and weak sites with adsorption
energy $\varepsilon^f_2$. As seen later, these sites can be
spatially distributed in different ways (different topographies).
Then, the total adsorption energy for an isolated $k$-mer on the
first layer with $k_1$ monomers located over strong sites and
$k_2$ monomers located over weak sites is
\begin{equation}
E_i =  k_1 \varepsilon^f_1 + k_2 \varepsilon^f_2. \label{etotal}
\end{equation}

Under these considerations, and using the formalism of the
integral equation of the adsorption isotherm \cite{Rudzinski1992},
the mean coverage $\theta$ can be written as
\begin{equation}
\theta =  \sum_{E_i} f(E_i) \theta_\mathrm{loc}(E_i), \label{isoh}
\end{equation}
where $f(E_i)$ is the fraction of $k$-uples of $k_1$ strong sites
and $k_2$ weak sites ($k_1+k_2=k$) and $\theta_\mathrm{loc}(E_i)$
represents the local multilayer adsorption isotherm corresponding
to an adsorptive energy $E_i$. This local isotherm can be well
approximated by using the multilayer adsorption isotherm
associated to an homogeneous surface characterized by an effective
value of $c$ given by
\begin{equation}
c_i=\exp \left[ - \beta \left( E_i - k \varepsilon \right)
\right].
\end{equation}
This value of $c_i$ can also be expressed as function of $c_1$ and
$c_2$, the values of $c$ for homogeneous surfaces whose adsorption
energies are $\varepsilon^f_1$ and $\varepsilon^f_2$,
respectively. Thus, if the $i$-th term in Eq. (\ref{isoh})
corresponds to a $k$-mer with $k_1$ units located over strong
sites and $k_2$ units located over weak sites, then
\begin{equation}
c_i=\left(c_1^{k_1} c_2^{k_2} \right)^{1/k}.
\end{equation}

As an example, let us consider the multilayer adsorption of
non-interacting dimers on a bivariate linear lattice, where strong
and weak sites are spatially distributed in alternating homotattic
patches of size $l$ ($l=1,2,3, \cdots$). In this case, Eq.
(\ref{isoh}) has three different terms, being each one of them a
dimer isotherm\footnote{In the particular case of $k=1$ and $k=2$,
it is possible to solve Eqs. (\ref{eq:7NTa}) and (\ref{eq:21NT})
to obtain single expressions of the multilayer isotherms
corresponding to monomers and dimers.} with a particular value of
$c$,
\begin{eqnarray}
\theta &=& \left( \frac{l-1}{2l} \right) \frac{1}{\left( 1- p/p_0
\right)} \left\{ 1 - \left[ \frac{1-p/p_0}{1+ \left( 4c_1-1
\right)p/p_0} \right]^{1/2}
\right\} + \nonumber\\
&+& \left( \frac{1}{l} \right) \frac{1}{\left( 1- p/p_0 \right)}
\left\{ 1 - \left[ \frac{1-p/p_0}{1+ \left( 4 \sqrt{c_1 c_2} -1
\right)p/p_0}
\right]^{1/2} \right\} + \nonumber\\
&+& \left( \frac{l-1}{2l} \right) \frac{1}{\left( 1- p/p_0
\right)} \left\{ 1 - \left[ \frac{1-p/p_0}{1+ \left( 4c_2-1
\right)p/p_0} \right]^{1/2} \right\}. \label{isoh1Dk2}
\end{eqnarray}
The first [third] term in the RHS of Eq. (\ref{isoh1Dk2})
represents the adsorption within a strong [weak] patch, on a pair
of sites ($\varepsilon^f_1,\varepsilon^f_1$)
[($\varepsilon^f_2,\varepsilon^f_2$)], with $c_1$ [$c_2$]. The
fraction of ($\varepsilon^f_1,\varepsilon^f_1$)
[($\varepsilon^f_2,\varepsilon^f_2$)] pairs on the lattice is
$(l-1)/2l$ \cite{PRE1}. The remaining term of Eq. (\ref{isoh1Dk2})
corresponds to a dimer isotherm on a
($\varepsilon^f_1,\varepsilon^f_2$) [or
($\varepsilon^f_2,\varepsilon^f_1$)] pair ($c=\sqrt{c_1 c_2}$),
being $1/l$ the fraction of this kind of pairs on the lattice. As
it can be observed, Eq. (\ref{isoh1Dk2}) depends on $l$ and the
dimer isotherm {\it sees} the topography.

In general, the number of terms in Eq. (\ref{isoh}) increases as
the adsorbate size $k$ is increased and this equation leads to a
complex solution. This scheme can be notoriously simplified
following the results in Ref. \cite{Sanchez2009}. In this paper,
the authors showed that:
\begin{itemize}
\item If $k \gg l$ (with $k>1$), the multilayer adsorption
isotherm can be represented by a single homogeneous isotherm
\begin{equation}
\theta = \theta_\mathrm{loc} \left( \sqrt{c_1 c_2} \right).
\label{l1}
\end{equation}

\item For a topography where $k \ll l$, the isotherm is
\begin{equation}
\theta = \frac{1}{2} \theta_\mathrm{loc} \left( c_1 \right)
+\frac{1}{2} \theta_\mathrm{loc} \left( c_2 \right). \label{LPT}
\end{equation}

\item The details of the topography are relevant only when $k \sim
l$. In this case, it is possible to consider a simpler expression
of the multilayer isotherm given by
\begin{equation}
\theta = \left( \frac{l-1}{2l} \right) \theta_\mathrm{loc} \left(
c_1 \right) + \left( \frac{1}{l} \right) \theta_\mathrm{loc}
\left( \sqrt{c_1 c_2} \right) + \left( \frac{l-1}{2l} \right)
\theta_\mathrm{loc} \left( c_2 \right). \label{isoh1Daprox}
\end{equation}
Eq. (\ref{isoh1Daprox}) $(i)$ captures the extreme behaviors Eqs.
(\ref{l1}) and (\ref{LPT}), and $(ii)$ approximates very well the
complete Eq. (\ref{isoh}) in 1D and 2D. In the 2D case, the local
isotherm in Eq. (\ref{isoh1Daprox}) is obtained from Eqs.
(\ref{eq:7NTa}) and (\ref{pre2D}), with $\zeta = 3$, $4$ and $6$
for honeycomb, square and triangular lattices, respectively.
\end{itemize}

Finally, we propose to extend Eq. (\ref{isoh1Daprox}) as to
describe multilayer adsorption of interacting polyatomic molecules
on heterogeneous surfaces. For this purpose, we propose to obtain
the local isotherms from Eqs. (\ref{eq:7NTa}) and (\ref{pqca}).
The advantages of using this simple description as a tool for
interpreting multilayer adsorption data and characterization of
the adsorption potential will be shown in Section 4 by analyzing
simulation results.

\section{Monte Carlo simulation}

The adsorption process is simulated through a grand canonical
ensemble Monte Carlo (GCEMC) method.

For a given value of the temperature $T$ and chemical potential
$\mu$, an initial configuration with $N$ $k$-mers adsorbed at
random positions (on $kN$ sites) is generated. Then, an
adsorption-desorption process is started, where each elementary
step is attempted with a probability given by the Metropolis
\cite{Metropolis1953} rule:
\begin{equation}
W =\min \left\{ 1, \exp{\left[- \beta \left( \Delta H - \mu \Delta
N  \right) \right]}\right\},
\end{equation}
where $\Delta H$ and $\Delta N$ represent the difference between
the Hamiltonians and the variation in the number of particles,
respectively, when the system changes from an initial state to a
final state. In the process there are four elementary ways to
perform a change of the system state, namely, adsorbing one
molecule onto the surface, desorbing one molecule from the
surface, adsorbing one molecule in the bulk liquid phase and
desorbing one molecule from the bulk liquid phase. In all cases,
$\Delta N= \pm 1$.

The algorithm to carry out one MC step (MCS), is the following :

\begin{itemize}
\item[1)]  Set the value of the chemical potential $\mu$ and the
temperature $T$.

\item[2)]  Set an initial state by adsorbing $N$ molecules in the
system. Each $k$-mer can adsorb in two different ways: $i)$ on a
linear array of ($k$) empty sites on the surface or $ii)$ exactly
onto an already adsorbed $k$-mer.

\item[3)]  Introduce an array, denoted as $A$, storing the
coordinates of $n_e$ entities, being $n_e$,
\begin{eqnarray}
n_e & = & {\rm number \ of \ available \ adsorbed \ {\it k}\rm{-mers} \ for \ desorption \ (n_d)} \nonumber \\
& + & {\rm \ number \ of \ available \ {\it k}\rm{-uples} \ for \
adsorption \ (n_a)},
\end{eqnarray}
where $n_a$ is the sum of two terms: $i)$ the number of $k$-uples
of empty sites on the surface and $ii)$ the number of columns of
adsorbed $k$-mers\footnote{Note that the top of each column is an
available $k$-uple for the adsorption of one $k$-mer.}.

\item[4)] Choose randomly one of the $n_e$ entities, and generate
a random number $\xi {\in }\left[ 0,1\right] $

\begin{itemize}
\item[4.1)]  if the selected entity is a $k$-uple of empty sites
on the surface then adsorb a $k$-mer if $\xi \leq W_{ads}^{surf}$,
being $W_{ads}^{surf}$ the transition probability of adsorbing one
molecule onto the surface.

\item[4.2)]  if the selected entity is a $k$-uple of empty sites
on the top of a column of height $i$, then adsorb a new $k$-mer in
the $i+1$ layer if $\xi \leq W_{ads}^{bulk}$, being
$W_{ads}^{bulk}$ the transition probability of adsorbing one
molecule in the bulk liquid phase.

\item[4.3)]  if the selected entity is a $k$-mer on the surface
then desorb the $k$-mer if $\xi \leq W_{des}^{surf}$, being
$W_{des}^{surf}$ the transition probability of desorbing one
molecule from the surface.

\item[4.4)]  if the selected entity is a $k$-mer on the top of a
column then desorb the $k$-mer if $\xi \leq W_{des}^{bulk}$, being
$W_{des}^{bulk}$ the transition probability of desorbing one
molecule from the bulk liquid phase.
\end{itemize}

\item[5)] If an adsorption (desorption) is accepted in $4)$, then,
the array $A$ is updated.

\item[6)]  Repeat from step $4)$ $M$ times.
\end{itemize}

In the present case, the equilibrium state could be well
reproduced after discarding the first $m\approx 10^6MCS$. Then,
averages were taken over $m^{\prime }\approx 10^6MCS$ successive
configurations. The total coverage was obtained as simple
averages,
\begin{equation}
\theta = {k\left< N \right> \over M},
\end{equation}
where $\left< N \right>$ is the mean number of adsorbed particles,
and $\left<...\right>$ means the time average over the MC
simulation runs.

\section{Results and discussion}

In the present section, we will analyze the main characteristics
of the multilayer adsorption isotherms given by Eq.
(\ref{isoh1Daprox}) [with local isotherms obtained from Eqs.
(\ref{eq:7NTa}) and (\ref{pqca})], in comparison with simulation
results for a lattice-gas of interacting $k$-mers on heterogeneous
one-dimensional and square lattices.

Heterogeneity is introduced by considering bivariate surfaces,
i.e., surfaces composed by two kinds of sites in the first layer,
strong and weak sites, with adsorptive energies $\varepsilon^f_1$
and $\varepsilon^f_2$, respectively. Recent developments in the
theory of adsorption on heterogeneous surfaces, like the supersite
approach \cite{Steele1999}, and experimental advances in the
tayloring of nanostructured adsorbates
\cite{Yang1998,Lopinski2000}, encourage this kind of study. A
special class of bivariate surfaces, with a chessboard structure,
has been observed recently to occur in a natural system
\cite{Fishlock2000}, although it was already intensively used in
modeling adsorption and surface diffusion phenomena
\cite{PRE1,SS8,PCCP1,SS13}.

Bivariate surfaces may also mimic, to a rough approximation, more
general heterogeneous adsorbates. Just to give a few examples, we
may mention the surfaces with energetic topography arising from a
continuous distribution of adsorptive energy with spatial
correlations, like those described by the dual site-bond model
\cite{Zgrablich1996}, or that arising from a solid where a small
amount of randomly distributed impurity (strongly adsorptive)
atoms are added \cite{Bulnes1999}. In both cases the energetic
topography could be roughly represented by a random spatial
distribution of irregular patches (with a characteristic size) of
weak and strong sites.

In the particular case studied in this article, the surface is
modeled in two different ways: $(1)$ as a chain of alternating
patches of size $l$ (see Fig. 2a); and $(2)$ as a collection of
finite homotattic patches in a chessboard-like array, where each
patch is assumed to be a domain of equal size, $l \times l$ sites
(see Fig. 2b). In this model, the energy correlation length is
simply given by the patch size.

The computational simulations have been developed for
one-dimensional chains of $10^4$ sites, and square $L \times L$
lattices with $L = 144$, and periodic boundary conditions. With
this lattice size we verified that finite-size effects are
negligible. Note, however, that the linear dimension $L$ has to be
properly chosen in such a way that it is a multiple of $l$.

We start analyzing what happens when the topography is changed.
Fig. 3 shows the behavior of the multilayer adsorption isotherms
for $k=2$, $\beta w=-1$ and different topographies in 1D as
indicated. The energy difference between different patches has
been chosen to be high ($c_1=1000$ and $c_2=1$) in order to
emphasize the effects of the surface heterogeneity. We have
identified the different topographies as $l_C$ for patches of size
$l$, and $bp$ for the case of a surface with two big patches ($l
\rightarrow \infty$). Symbols represent simulation results and
lines correspond to theoretical data [Eq. (\ref{isoh1Daprox})]. It
can be seen that all curves are contained between the two limit
ones: the one corresponding to $1_C$ and the one corresponding to
$bp$.

For $l=1$, the adsorption energy of a dimer in the first layer is
$\varepsilon^f_1 + \varepsilon^f_2$ for all configuration. In this
condition, the system corresponds to a 1D lattice-gas of
interacting dimers on a homogeneous surface and, consequently, Eq.
(\ref{isoh1Daprox}) is exact.

In general, for $l>1$, this equation is approximate. For $l=2$,
the analytic isotherms agree very well with the simulation data.
However, for $p/p_0$ ranging between 0 and 0.15, some differences
between theoretical and numerical data are observed. This happens
because Eq. (\ref{isoh1Daprox}) has been built assuming that the
three different pairs of sites are filled simultaneously and
independently.  However, for $c_1 \gg c_2$, the real process
occurs in 3 stages: $(i)$ the pairs of sites
($\varepsilon^f_1,\varepsilon^f_1$) are covered; $(ii)$ the pairs
($\varepsilon^f_2,\varepsilon^f_2$) begin to be filled and $(iii)$
the multilayer is formed. Note that in the first stage all the
pair of sites ($\varepsilon^f_1,\varepsilon^f_2$) and
($\varepsilon^f_2,\varepsilon^f_1$) are removed.  For this regime,
a better approximation can be obtained by a semisum of two
isotherms with $c_1$ y $c_2$.

When $l=3$, the agreement between the analytic isotherms and the
simulation data is very good. In this case, for $c_1 \gg c_2$ the
first stage does not eliminate all the pairs of sites
($\varepsilon^f_1,\varepsilon^f_2$) and
($\varepsilon^f_2,\varepsilon^f_1$), because each dimer occupies
only two sites in the strong patches. For this reason, the range
of validity of Eq. (\ref{isoh1Daprox}) is wider than in the
previous case.  Now, if $l=4$ or $l=5$, the behaviors are similar
to those observed for $l=2$ or $l=3$, respectively. In general,
for even $l$, the first stage eliminates almost completely the
pairs of sites ($\varepsilon^f_1,\varepsilon^f_2$) and
($\varepsilon^f_2,\varepsilon^f_1$), while this does not happen
for odd $l$. Finally, when $l \to \infty $, the fraction of pair
($\varepsilon^f_1,\varepsilon^f_2$) and
($\varepsilon^f_2,\varepsilon^f_1$) goes to zero and Eq.
(\ref{isoh1Daprox}) is exact. This limit corresponds to the called
large patches topography ($bp$ surface in our model), where the
surface is assumed to be a collection of homogeneous patches,
large enough to neglect border effects between neighbor patches
with different adsorption energies.

We now analyze the effect of the lateral interactions on the
behavior of the system. For this purpose, Fig. 4 shows the
adsorption isotherms for $k=2$, $c_1=1000$, $c_2=1$ and two
different values of the lateral interactions $\beta w=-1$
(attractive case) and $\beta w=1$ (repulsive case). In addition,
for each value of $\beta w$, the limit topographies ($1_C$ and
$bp$) have been considered (as seen in Fig. 3, all curves
corresponding to all topographies are contained between them).

For repulsive couplings, the interactions do not favor the
adsorption on the first layer and the isotherms shift to higher
values of pressure. On the other hand, attractive lateral
interactions facilitate the formation of the monolayer.
Consequently, the isotherms shift to lower values of $p/p_o$ and
their slope increases as the ratio $|\beta w|$ increases. In both
cases ($\beta w=-1$ and $\beta w=1$), the agreement between
theoretical and simulation data is excellent.

From the curves in Fig. 4 (and from data not shown here for the
sake of clarity) it is observed that: there exists a wide range of
$\beta w$'s ($-2 \leq \beta w \leq 2$), where the theory provides
an excellent fitting of the simulation data. In addition, most of
the experiments in surface science are carried out in this range
of interaction energy. Then, the present theory not only
represents a qualitative advance in the description of the
multilayer adsorption of interacting $k$-mers on heterogeneous
surfaces, but also gives a framework and compact equations to
consistently interpret thermodynamic multilayer adsorption
experiments of polyatomics species such as alkanes, alkenes, and
other hydrocarbons on regular surfaces.

The effect of energetic heterogeneity (ratio between $c_1$ and
$c_2$) is analyzed in Fig. 5, where the degree of heterogeneity is
varied by changing the value of $c_2$ between 1 and 100 with $c_1$
fixed ($c_1=1000$). As in Fig. 4, $k=2$ and $\beta w=-1$. Lines
represent theoretical results and symbols correspond to simulation
data ($c_2=1$: circles; $c_2=10$: triangles; and $c_2=100$:
squares). For each set of values of the parameters, the limit
cases corresponding to $1_C$ (open symbols) and $bp$ (solid
symbols) topographies are studied. As it can be observed from the
simple inspection of the figure, the effect of topography is
important in a range of $c_2/c_1$ between $10^{-3}$ and $10^{-2}$
and is practically negligible for $10^{-1} < c_2/c_1 <1 $.

To complete the discussion started in Fig. 3, we now evaluate the
effect of the $k$-mer size on the adsorption isotherms. This study
is shown in Fig. 6, where the $1_C$ and $bp$ multilayer adsorption
isotherms are plotted for $\beta w=-1$, $c_1=1000$, $c_2=1$ and
two different values of $k$ ($k=2$ and $k=10$). One important
conclusion can be drawn from the figure. Namely, the effects of
topography and energetic heterogeneity tends to disappear as the
size $k$ is increased.

The study in Figs. 3-6 was repeated for surfaces in 2D. The
behavior of the curves (not shown here for brevity) is very
similar to the one observed in 1D.

Summarizing, we have shown that just by using an expression of
three terms, Eq. (\ref{isoh1Daprox}), we can approach very well
the multilayer isotherm in 1D and 2D for the adsorption of
interacting polyatomics on heterogeneous surfaces. In the next, we
will use this approximation and Monte Carlo simulations to study
how lateral interactions, multisite occupancy and surface
heterogeneity affect  the determination of monolayer volume
predicted by the BET equation.

In a typical experiment of adsorption, the adsorbed volume of the
gas, $v$, is measured at different pressures and at a given fixed
temperature, the total coverage is $\theta=v/v_\mathrm{m}$.
Analyzing an isotherm with the BET equation
\begin{equation}
\theta = \left( \frac{1}{1-p/p_0} \right)  \left( \frac{c
p/p_0}{1-p/p_o+cp/p_0} \right),
\end{equation}
it is possible to estimate the monolayer volume if we rewrite the
previous equation as:
\begin{equation}
\frac{p/p_0}{v \left( 1- p/p_0 \right)}=\frac{1}{c
v_\mathrm{m}}+\frac{\left( c-1 \right)}{c v_\mathrm{m}} p/p_0 .
\label{BETlin}
\end{equation}
This equation is a linear function of $p/p_0$.  If we denote with
$a$ and $b$, the $y$-intercept and the slope of this straight
line, respectively, we obtain
\begin{equation}
v_\mathrm{m}^*=\frac{1}{a+b} \label{vas}
\end{equation}
and
\begin{equation}
c^*=\frac{b}{a}+1.\label{cas}
\end{equation}
The asterisk has been added in order to indicate that the
quantities given by Eqs. (\ref{vas}) and (\ref{cas}) correspond to
the prediction of the BET theory. Then, by means of a plot (the
so-called BET plot) of the experimental data of $\frac{p/p_0}{v
\left( 1- p/p_0 \right)}$ vs $p/p_0$, we can obtain an estimate of
the monolayer volume and the parameter $c$. Nevertheless, in the
experiments it is commonly found that there are deviations from
linearity in the BET plot.

Following the scheme described in previous paragraphs, we carry
out numerical experiments to determine, in different adsorption
situations, how much the value of the monolayer volume predicted
by the BET equation differs from its real value, $v_\mathrm{m}$.
With this purpose, analytic and simulation isotherms were analyzed
as experimental data. In this way, we have determined how
adsorbate size, surface heterogeneity and lateral interactions,
affect the standard determination of the monolayer volume.

In Fig. 7 we plot the calculated value of
$v_\mathrm{m}^*/v_\mathrm{m}$ as a function of $k$ for fixed
values of $c_1$ and $c_2$ (namely $c_1=1000$ and $c_2=1$) and
different values of the attractive lateral interaction. The
repulsive interactions, not shown in the figure, show a marked
deviation from the predicted BET equation, therefore wiping out
compensation effects due to the surface heterogeneity.

We can see three sets of plots in Fig. 7. Each set (1) is the
result of the effect of different lateral interactions strengths:
$\beta w=0$, circles; $\beta w=-0.5$, squares and $\beta w=-1.0$,
triangles; and (2) depicts the limiting cases for the surface
topography: open and solid symbols correspond to $1_C$ and $bp$
surfaces, respectively. All other topographies must lay in between
those two plots.

If we analyze the behavior of the open symbols, we are analyzing,
essentially, the homogenous case \cite{Sanchez2009} where an
already known feature for the homogeneous scenario can be
distinguished: the compensation effect of $k$ is lower as $k$
increases, as \cite{Roma2005} stated for the non-interacting case.
On the other hand, for $bp$ surfaces, the compensation effect
increases with $k$ as seen in \cite{Sanchez2009}. In all cases,
for stronger (attractive) interaction strength the plots move
upwards showing greater compensation effects.

As shown in previous work \cite{Riccardo2002,Roma2005} for
non-interacting $k$-mers on homogeneous surfaces, the monolayer
volume from the BET model diminishes with increasing values of the
$k$-mer size. The data presented in Fig. 7 demonstrate that
attractive lateral interactions and surface heterogeneity play a
key role in the compensation of $k$-mer size effects. This finding
is very important because most of the experiments in surface
science are carried out in these conditions.

Finally, the study in Fig. 7 is repeated for 2D surfaces (see Fig.
8). In this case, both the compensation effect for attractive
interactions and the underestimation of the monolayer volume for
repulsive interactions (not shown here) are more important than in
the 1D case. The explanation is simple: in 1D systems, particles
will interact only at their ends, regardless of $k$; on the other
hand, in 2D systems, each particle interacts at their ends, but
also interacts along its $k$ monomers. This makes the interaction
energy grow linearly with $k$ (if all first neighbors are
occupied), and indicates that lateral interactions will play a
more important role in two-dimensional adsorption systems than in
one-dimensional ones.

\section{Conclusions}

In this work, we have studied the multilayer adsorption of
interacting polyatomic molecules onto heterogeneous surface. The
polyatomic character of the adsorbate was modeled by a lattice gas
of $k$-mers. With respect to lateral interactions, the ad-ad
couplings in the monolayer were explicitly considered in the
solutions. The range of validity of analytical isotherms was
analyzed by comparing theoretical and MC simulation results.

We also analyzed the 1D and 2D BET plots obtained using the
theoretical and simulation isotherms. For non-interacting
$k$-mers, we found that the use of BET equation leads to an
underestimate of the true monolayer volume: this volume diminishes
as $k$ is increased. The situation is different for the case of
interacting molecules over heterogeneous surface. Thus, attractive
lateral interactions favor the formation of the monolayer and,
consequently, compensate the effect of the multisite occupancy. In
this case, the monolayer volume predicted by BET equation agrees
very well with the corresponding true value. In the case of
repulsive couplings, the lateral interactions impede the formation
of the monolayer and the BET predictions are bad (even worse than
those obtained in the non-interacting case). Both the compensation
effect for attractive interactions and the underestimation of the
monolayer volume for repulsive interactions are more important for
2D systems.

\section{ACKNOWLEDGMENTS}

This work was supported in part by CONICET (Argentina) under
project number PIP 112-200801-01332; Universidad Nacional de San
Luis (Argentina) under project 322000; Universidad Tecnol\'ogica
Nacional, Facultad Regional San Rafael (Argentina) under projects
PQPRSR 858 and PQCOSR 526 and the National Agency of Scientific
and Technological Promotion (Argentina) under project 33328 PICT
2005.

\newpage

\section*{Figure Captions}

\noindent Fig. 1: Schematic representation of the evolution of
complexity in the theoretical adsorption models.

\noindent Fig. 2: Schematic representation of heterogeneous
bivariate square surfaces with chessboard topography. The black
(white) symbols correspond to strong (weak) adsorption sites. (a)
One-dimensional lattice and (b) square lattice. The patch size in
this figure is $l = 2$.

\noindent Fig. 3: Adsorption isotherms for dimers on
one-dimensional lattices with $\beta w=-1$ and different
topographies 1D as indicated. The energy difference between
different patches has been chosen to be high ($c_1=1000$ and
$c_2=1$) in order to emphasize the effects of the surface
heterogeneity. Solid lines and symbols represent theoretical and
simulation results, respectively. For each set of values of the
parameters, the limit cases corresponding to $1_C$ (open symbols)
and $bp$ (solid symbols) topographies are shown.

\noindent Fig. 4: Adsorption isotherms for dimers on
one-dimensional lattices with $c_1=1000$, $c_2=1$ and two
different values of the lateral interactions $\beta w=-1$
(attractive case) and $\beta w=1$ (repulsive case). Solid lines
and symbols represent theoretical and simulation results,
respectively. For each set of values of the parameters, the limit
cases corresponding to $1_C$ (open symbols) and $bp$ (solid
symbols) topographies are shown.

\noindent Fig. 5: Adsorption isotherms for dimers on
one-dimensional lattices with $\beta w=-1$. The degree of the
surface heterogeneity is varied by changing the value of $c_2$
between 1 and 100 with $c_1$ fixed ($c_1=1000$). Lines represent
theoretical results and symbols correspond to simulation data
($c_2=1$: circles; $c_2=10$: triangles; and $c_2=100$: squares).
For each set of values of the parameters, the limit cases
corresponding to $1_C$ (open symbols) and $bp$ (solid symbols)
topographies are shown.

\noindent Fig. 6: Adsorption isotherms for dimers on
one-dimensional lattices with $\beta w=-1$, $c_1=1000$, $c_2=1$
and two different values of $k$ ($k=2$ and $k=10$). Solid lines
and symbols represent theoretical and simulation results,
respectively. For each set of values of the parameters, the limit
cases corresponding to $1_C$ (open symbols) and $bp$ (solid
symbols) topographies are shown.

\noindent Fig. 7: Results of the BET plots for the adsorption in
1D heterogeneous surfaces with $c_1=1000$, $c_2=1$. Dependence on
$k$ of the fraction $v_\mathrm{m}^* / v_\mathrm{m}$ for three
different values of $\beta w$: $\beta w=0$, circles; $\beta
w=-0.5$, squares and $\beta w=-1.0$, triangles. For each set of
values of the parameters, the limit cases corresponding to $1_C$
(open symbols) and $bp$ (solid symbols) topographies are shown.

\noindent Fig. 8: As Fig. 7 for 2D surfaces: $\beta w=0$, circles;
$\beta w=-0.1$, squares and $\beta w=-0.5$, triangles. Symbols
connected by (solid) dotted lines correspond to results obtained
from (theoretical) MC simulation isotherms.

\end{document}